\documentclass[12pt]{article}

\usepackage[paper=a4paper, margin=2cm]{geometry}
\usepackage[russian,english]{babel}
\usepackage[cp1251]{inputenc}
\usepackage{amsmath}

\usepackage[logos]{amsbib}

\begin{document}

\date{}
\renewcommand{\refname}{References}

\author{A.~Yu.~Samarin
}
\title{Nonlinear dynamics of open quantum systems}

\maketitle

\centerline{Samara Technical State University, 443100 Samara, Russia}
\centerline{}

\abstract{ The evolution of a composite closed system using the integral wave equation with the kernel in the form of path integral is considered. It is supposed that a quantum particle is a subsystem of this system. The evolution of the reduced density matrix of the subsystem is described on the basis of the integral wave equation for a composite closed system. The equation for the density matrix for such a system is derived. This equation is nonlinear and depends on the history of the processes in the closed system. It is shown that, in general, the reduced density matrix trace does not conserve in the evolution processes progressing in open systems and the procedure of the trace  normalization is necessary as the mathematical image of a real nonlocal physical process. The wave function collapse and EPR correlation are described using this approach.



\section {introduction}

 The principle differences between the dynamics of an open system and the evolution of a closed one can not solely restricted by the irreversibility of the former. In general, there exist nonlinear transformations of the reduced density matrices of open systems.  Depending on the specific properties of the system, these nonlinear processes can take the form of the wave function collapse in process of the measurement, the decoherence phenomenon, etc. Any open system can be considered as a subsystem of a large closed system obeying the linear evolution law. The impossibility to describe the nonlinear state transformation of an open system under the measurement using the Schr\"odinger equation led to the necessity to formulate a particular reduction postulate~\cite{bib:1} (the quantum jump notion~\cite{bib:2}). The peculiarity of the problem is that there is no cause, expressed in precise physical terms, determining the form of the transformation of the quantum state~\cite{bib:3}.
 
 Except for nonlinearity the open system dynamics, in general, has one more specific property --- the dependence on the evolution history. This property already emerges in the correlation of the uncertainty of the measured value of the stationary state energy with the duration of the measurement process and becomes apparent when considering the EPR paradox~\cite{bib:4}\footnote{The future states of entangled subsystems are not completely determined by their reduced density matrices at the current time.}. 
 
 We assume that the Schr\"odinger equation is absolutely accurate when describing the evolution of closed quantum systems for infinitesimal time intervals\footnote{As opposed, for example, to the paper~\cite{bib:5}, we do not modify Schr\"odinger's equation and introduce no terms in it.}. The unique strict generalization of Schr\"odinger's equation on finite time intervals is the integral wave equation with the kernel in the form of path integral~\cite{bib:6,bib:7}. The action functionals entering into the integral evolution operator generates the dependence of the quantum system state on the evolution history. Besides, the mathematical form of this law supposes the existence of a subsystem nonlinear evolution~\cite{bib:8,bib:9}.    

 Open system quantum states can be described by reduced density matrices. A corresponding evolution equation is usually derived by considering a large closed system including this open system S and the rest part of the closed system R. It is supposed that the evolution of the closed system is Schrodinger's one. The next assumption is that the state of the rest part R of the closed system is not affected by the state of the subsystem S and the open system evolution can be considered as a Marcov process\footnote{Assuming that the Marcov process happens in the open system, the state of the subsystem R is described statistically, which does not allow us deterministically  consider the closed system quantum dynamics. The approach offered here uses the deterministic equation only, so that the terms the Markov process and the master equation are not used here.}(Born-Markov approximation)~\cite{bib:10,bib:11}.
	
Keeping the model of an open system as a part of a closed system we consider the closed system in a pure quantum state and, therefore, the evolution processes of this system, as well as those of all its subsystems are deterministic. The processes in the subsystems depend on the evolution history of the closed system and they are described exactly by the corresponding integral wave equation with the kernel in the form of path integral. Following this approach, the deterministic equation for the reduced density matrix of the subsystem S is derived.
 
  \section {The equation for the subsystem's density matrix }
 
 Let a quantum particle be the considered subsystem S. We shall limit ourselves to an one-dimensional problem as the generalization to several dimensions is obvious. The closed system wave function $\Psi_t$ is the function of the particle position (denote it by $x$ or $y$) and the generalized coordinates of the rest part of the closed system (denote by $q$ the set of them, by $dq=dq_1\cdot\cdot\cdot dq_s$ --- the configuration space volume, where $s$ is the number of the subsystem degrees of freedom ). The reduced density matrix of the particle at the time $t$ is
\begin{equation*}
 \rho_t(x,y)=\int\limits_{-\infty}^{\infty}\Psi_t^*(y,q)\Psi_t(x,q)dq,
\end{equation*}
Suppose that the subsystems (the particle and the rest part of the system) do not interact before the time $t'$. Then, for the evolution of the closed system state after the time $t'$, we have
\begin{equation*}
 \Psi_{t}(x,q)=\int\limits_{-\infty}^{\infty}\int\limits_{-\infty}^{\infty} K_{t,t'}(x,q,x',q')\Psi_{t'}(x')\Phi_{t'}(q')\,dx'dq',
\end{equation*}
where $\Psi_{t'}(x')$ and $\Phi_{t'}(q')$ are the wave functions of the subsystems before the interaction; the transition amplitude $K_{t,t'}(x,q,x',q')$ is described by the continual integral~\cite{bib:12}
\begin{equation*}
 K_{t,t'}(x,q,x',q')=\int\int \exp\frac{i}{\hbar}\Bigl(S_1[x(t)]+S_2[q(t)]-I[x(t),q(t)]\Bigr)\,[dx(t)][dq(t)]; 
\end{equation*}
$S_1[x(t)]=\int\limits_{t'}^{t}\bigl(T_1(\dot{x})-U_1(x)\bigr)\,dt$, $S_2[q(t)]=\int\limits_{t'}^{t}\bigl(T_2(\dot{q})-U_2(q)\bigr)\,dt$ are respectively the actions for the virtual paths of the particle and the rest part of the closed system without interaction ($T$ and $U$ are kinetic and potential energies of the subsystems); $I[x(t),q(t)]=\int\limits_{t'}^{t''}V(x,q)\,dt$ is the functional describing the subsystems interaction ($V(x,q)$ --- the interaction energy of the subsystems; $t''$  --- the time of the interaction termination if $t''\leq t$, if not, then $t''=t$). Then, the reduced density matrix at the time $t>t'$ is 
\begin{alignat}{2}\label{eq:math:ex1}
 \rho_t(x,y)=\int\limits_{-\infty}^{\infty}\int\limits_{-\infty}^{\infty}\Biggl(\int\int\exp\biggl(-\frac{i}{\hbar}S_1[y(t)]\biggr)\exp\frac{i}{\hbar}S_1[x(t)]\times\nonumber\\\times\exp\biggl(- S^A[x(t),y(t)]\biggr)\,[dy(t)][dx(t)]\Biggr)\Psi_{t'}^*(y')\Psi_{t'}(x')\,dy'dx'.
\end{alignat}
The functional $S^A[x(t),y(t)]$, associating the subsystems, describes the influence of the environment on the particle evolution. It has the form:
\begin{alignat}{2}\label{eq:math:ex2}
 S^A[x(t),y(t)]=\nonumber\\=-\ln\int\limits_{-\infty}^{\infty}\Biggl(\biggl(\int\limits_{-\infty}^{\infty}\Bigl(\int\exp\frac{i}{\hbar}(-S_2[q(t)]+I[y(t),q(t)]\bigr)\,[dq(t)]\Bigr)\Psi_{t'}^*(q')dq'\biggr)\times\nonumber\\\times\biggl(\int\limits_{-\infty}^{\infty}\Bigl(\int\exp\frac{i}{\hbar}(S_2[q(t)]-I[x(t),q(t)]\bigr)[dq(t)]\Bigr)\Psi_{t}'^(q')dq'\biggr)\Biggr)\,dq.
\end{alignat}
To convert equation~\eqref{eq:math:ex1} to a differential form, let us consider the development of the density matrix with time for the shot finite interval of time $\varepsilon$. Denote by $\eta$ and $\zeta$ the increments of arguments $x(t+\varepsilon)-x(t)$ and $y(t+\varepsilon)-y(t)$ respectively. Then, using the results of the book~\cite{bib:12}, the evolution of the subsystem density matrix for the time interval $\varepsilon$ is governed by the equation:
\begin{alignat*}{2}
\rho_{t+\varepsilon}(x,y)=\\=\int\limits_{-\infty}^{\infty}\exp\biggl(-\frac{i}{\hbar}\Bigl(\frac{m\zeta^2}{2\varepsilon}-U_1\Bigr)\biggr)\Biggl(\int\limits_{-\infty}^{\infty}\exp\frac{i}{\hbar}\Bigl(\frac{m\eta^2}{2\varepsilon}-U_1\varepsilon-\frac{\hbar}{i}\frac{\partial S^A}{\partial t}\varepsilon\Bigr)\rho_t(x-\eta,y-\zeta)\,d\eta \Biggr)d\zeta.
\end{alignat*}
To transform the integral into the real form, we replace the real time $t$ with the imaginary one $t=i\tau$ (or  $t=-i\tau$ for the complex conjugated transition amplitude), where $\tau$ is a complex time modulus . Expanding all the terms of the last equation as a Taylor series to the first order of smallness and equating the terms of the first order in $\varepsilon$, for imaginary time, we obtain
\begin{equation*}
 \hbar\frac{\partial\rho_{\tau}(x,y)}{\partial \tau}=\frac{\hbar^2}{2m}\biggl(\frac{\partial^2}{\partial y^2}-\frac{\partial^2}{\partial x^2}\Biggr)\rho_{\tau}(x,y)+\biggl(U(x)-U(y)+\hbar\frac{\partial S^A(x,y)}{\partial\tau}\biggr)\rho_{\tau}(x,y).
\end{equation*} 
Analytic continuation to real time transforms this equation into the form:
\begin{equation}\label{eq:math:ex3}
 i\hbar\frac{\partial\rho_{t}(x,y)}{\partial t}=\frac{\hbar^2}{2m}\biggl(\frac{\partial^2}{\partial y^2}-\frac{\partial^2}{\partial x^2}\Biggr)\rho_{t}(x,y)+\biggl(U(x)-U(y)+i\hbar\frac{\partial S^A(x,y)}{\partial t}\biggr)\rho_{t}(x,y).
\end{equation}
Equation~\eqref{eq:math:ex3} describes the evolution of the subsystem deterministically. There are no stochastic terms in it. This equation can be nonlinear if the last term on the  right side of the equation is not zero. The form of this dependence is determined by the concrete situation of the interaction. Consider some examples of interaction types.
 
\section {Measurement of the particle position}\label{1}

Suppose that the localized in space macroscopic process (registering process) is initiated in the apparatus R as the result of interaction with the quantum particle S. The coordinate measuring instrument contains many elements where such processes can be initiated independently each other. Let the size of a single element $j$ ($j=\overline{1,N}$) be small enough the particle object wave function to be considered as having the same values in the volume of a single element. This assumption allows for describing the interaction of the particles of the apparatus element $j$ (active particles) with the particle-object by the unique coordinate $X_j$ of the mass center of these particles~\cite{bib:13,bib:14}. Then, the wave function of the system has the form:
\begin{equation*}
 \Psi_t(x,X_1,...,X_N)=\idotsint\prod\limits_{j=1}^{N}K(x,X_j,x',X'_j)\Psi(x',X'_j)\,dx'dX'_j,
\end{equation*}
and for functional~\eqref{eq:math:ex2}, we have
\begin{equation*}
 S^A=\sum\limits_{j=1}^{N}S^A_j,
\end{equation*}
where
\begin{alignat*}{2}
 S_j^A[x(t),y(t)]=\nonumber\\=-\ln\int\limits_{-\infty}^{\infty}\Biggl(\biggl(\int\limits_{-\infty}^{\infty}\Bigl(\int\exp\frac{i}{\hbar}(-S_2[X_j(t)]+I[y(t),X_j(t)]\bigr)\,[dX_j(t)]\Bigr)\Psi_{t'}^*(X'_j)\,dX'_j\biggr)\times\nonumber\\\times\biggl(\int\limits_{-\infty}^{\infty}\Bigl(\int\exp\frac{i}{\hbar}(S_2[X_j(t)]-I[x(t),X_j(t)]\bigr)[dX_j(t)]\Bigr)\Psi_{t}'^*(X'_j)\,dX'_j\biggr)\Biggr)\,dX_j.
\end{alignat*}
 Let, at first, the registering process be initiated in the element $k$ (at the time $t_r$). Denote by $U^A$ the potential energy of the active particles taking part in the registering process. It has a macroscopic value and after a very small time interval $\varepsilon$ the functional $\int\limits_{t_r}^{t_R+\varepsilon}U^A\,d\tau$ strongly exceeds any microscopic action in the transition amplitudes $K(x,X_j,x',X'_j)$. At the time $t_r+\varepsilon$, this functional equals approximately $U^A\varepsilon$ and for $S^A_k$, we have 
\begin{alignat*}{2}
 S_k^A[x(t),y(t)]=\nonumber\\=-\ln\int\limits_{-\infty}^{\infty}\Biggl(\biggl(\int\limits_{-\infty}^{\infty}\Bigl(\int\exp\frac{i}{\hbar}(-S_2[X_k(t)]+I[y(t),X_k(t)]+U^A\varepsilon\bigr)\,[dX_j(t)]\Bigr)\Psi_{t'}^*(X'_k)\,dX'_k\biggr)\times\nonumber\\\times\biggl(\int\limits_{-\infty}^{\infty}\Bigl(\int\exp\frac{i}{\hbar}(S_2[X_k(t)]-I[x(t),X_k(t)]-U^A\varepsilon\bigr)[dX_k(t)]\Bigr)\Psi_{t'}(X'_k)\,dX'_k\biggr)\Biggr)\,dX_k.
\end{alignat*} 
The actions in the last expression cannot be neglected in comparison with the quantity $U^A\varepsilon$, because they generate different phases for different virtual paths and the path integrals cannot be canceled. The modulus of continual integral is determined by the measure of the set of the paths defined by their kinetic energies and the weight of each path is defined by its potential energy. Since the set of the paths $X_j(t)$ at the time $t_r$ are approximately the same for all $S^A_j$ (as well as the modulus of the initial wave functions $\Psi_{t_r}(X_j)$), the orders of $S^A_j$   are determined by the corresponding potential energies which are the same, too, excepting for the element $k$ that has the macroscopic energy $U^A$ in the action after the time $t_r$. In order to estimate the consequences of this, it is necessary to transform the path integrals in the expressions for $S_K$ and $S_j$ into a real form. According to the conventional method we express the real time in the complex form $t=\tau\exp i\varphi$, but we take $\varphi=\frac{\pi}{2}$  (and not $\varphi=-\frac{\pi}{2}$ as in a usual case, for example in~\cite{bib:12})\footnote{The change $t$ for $-i\tau$ reverses the time, what is inadmissible for irreversible processes. Although the path integral measure (the Wiener measure) is defined for the imaginary negative time, it can be analytically continued on the upper part of the complex plate~\cite{bib:15}}. For the complex conjugated path integral, in consequence of reversed time flow, we have to take $\varphi=-\frac{\pi}{2}$. Taking into account the macroscopic value $U^A$  (and that $S^A_j>0$), we have 
\begin{equation*}
 S^A_k\sim\exp\frac{2U^A\varepsilon}{\hbar}S^A_j,
\end{equation*}
and, therefore,
\begin{equation*}
 \frac{\partial S^A_k}{\partial t}>>\sum\limits_{\substack{j=1\\j\neq k}}^N\frac{\partial S^A_j}{\partial t}
\end{equation*}
Besides, this term considerably exceeds all the other terms on the right side of equation~\eqref{eq:math:ex3} which takes the form:
\begin{equation*}
 \frac{\partial\rho(x,y)}{\partial t}=\frac{\partial S^A_k(x,y)}{\partial t}\rho(x,y).
\end{equation*}
Suppose that the interaction radius is much less than the apparatus active element size and that the apparatus elements sizes is infinitesimal. Then, setting the derivative $\frac{\partial S^A_k(x,y)}{\partial t}$ infinite large, after normalization of the density matrix trace, we have
\begin{equation*}
\rho(x,y)=\delta(x-X_k)\delta(y-X_k).
\end{equation*}

Thus, the wave function collapse is the result of the specific deterministic process\footnote{In quite a few publications the possibility of the deterministic collapse nature is considered as contradictory to the special relativity theory. In the conclusion we will show that the deterministic collapse mechanism offered here cannot contradict to it, in principle.}. Since the apparatus elements are macroscopic objects, they have different properties with respect to the interaction with quantum objects. Hereupon, the probabilities of the registering process initiation $p_j(\rho)$ are different for these elements and the quantum objects ensemble after measurement is a statistical mixture described by a diagonal density matrix. As opposite to the decoherence mechanism this diagonalization is the result of the collapse of the single quantum object state into a unique eigenstate after the measurement, and not the suppression of the transitions between different eigenstates of this object (such a process is considered in the paper~\cite{bib:5}). In the offered approach, the randomness of the wave function collapse is the result of the statistical straggling of physical properties of the macroscopic apparatus active elements, and, therefore, the probabilities have a epistemic character, i.e. they are due to our ignorance about the precise state of the macroscopic apparatus. 
 
\section {EPR paradox}

Let us suppose that the closed system consists of two quantum particles interacting in the past. Denote by $x_1$ (or $y_1$) and $x_2$ the positions of the particles in space (as before, we consider a one-dimensional motion). Let the subscript $1$ denote the quantities of the subsystems $S$ and $2$ --- $R$. Suppose that the interaction is elastic collision at the time $t'=t''$, then the particles move freely (before this time the particles do not interact). The functional $I(x_1,x_2)$ entangling the states of the particles in expression~\eqref{eq:math:ex2} has the same values for any virtual path of the system corresponding to this collision. It can be omitted while the entanglement of the particle states is conserved as the result of correlation between the paths of the sets $\{x_1(t)\}$ and $\{x_2(t)\}$ forming the set of the virtual paths of the system:
\begin{equation}\label{eq:math:ex4}
\begin{aligned}
p_1&=-p_2=p,\\
x''_1&=x''_2,\\
x_1&=x''_1+\frac{p}{m_1}(t-t''),\\
x_2&=x''_2-\frac{p}{m_2}(t-t''),
\end{aligned}
\end{equation}
where $p_1$ and $p_2$ are the particles momenta, $p$ is the absolute value of these momenta. Supposing that the wave functions of the particles immediately after the collision have the form of the Dirac delta-functions, for the functional $S^A$, we have
\begin{alignat*}{2}
 S^A[x(t),y(t)]=-\ln\frac{1}{(2\pi\hbar)^2}\times\nonumber\\\times\int\limits_{-\infty}^{\infty}dx_2\Biggl(\int\limits_{-\infty}^{\infty}\biggl(\int\exp\biggl(\frac{i}{\hbar}\int\limits_{t''}^{t}\frac{m\dot{x_2}^2}{2}\,dt\biggr)\,[dx_2(t)]\int\limits_{-\infty}^{\infty}\exp\Bigl(-\frac{i}{\hbar}p(x''_2+x_0)\Bigr)dp\biggr)\biggr|_{x_1(t)}dx''_2&\times\nonumber\\\times\int\limits_{-\infty}^{\infty}\biggl(\int\exp\biggl(-\frac{i}{\hbar}\int\limits_{t''}^{t}\frac{m\dot{x_2}^2}{2}\,dt\biggr)\,[dx_2(t)]\int\limits_{-\infty}^{\infty}\exp\Bigl(\frac{i}{\hbar}p(x''_2+x_0)\Bigr)\,dp\biggr)\biggr|_{y_1(t)}dx''_2&\Biggr)
\end{alignat*}
The subscriptions $x(t)$ and $y(t)$ specify the paths of the subsystem $S$, corresponding to the transition amplitude of the subsystem $R$. After integrating over the paths of free motion ~\cite{bib:7,bib:12}, the functional $S^A$ takes on the form:   
\begin{alignat*}{2}
 S^A[x(t),y(t)]=\nonumber\\=-\ln\frac{1}{(2\pi\hbar)^2}\int\limits_{-\infty}^{\infty}\Biggl(\int\limits_{-\infty}^{\infty}\biggl(\exp\frac{i}{\hbar}\Bigl(-p(x_2+x_0)-\frac{p^2}{2m_2}(t-t'')\Bigr)\biggr)\biggr|_{x_1(t)}\,dp\times\nonumber\\\times\int\limits_{-\infty}^{\infty}\biggl(\exp\frac{i}{\hbar}\Bigl(p(x_2+x_0)+\frac{p^2}{2m_2}(t-t'')\Bigr)\biggr)\biggr|_{y_1(t)}\,dp\Biggr)\,dx_2.
\end{alignat*}
If the momentum of the subsystem $S$ is specified, then, in accordance with the relations~\eqref{eq:math:ex4}, the paths $x_1(x''_1,t)$ and $y_1(y'',t)$ are well defined. Thus, we have the quantity $S^A$ in the form of a functional on the sets of the virtual paths of the system $S$.

If the value $x_{2m}$ of the position of the particle $R$ is obtained after the measurement at the time $t_r$, then the factor
\begin{equation*}
\exp\frac{i}{\hbar}U^A(t-t_r)
\end{equation*}
appears in the transition amplitude formed by the paths having origin in the position $x_0$ at the time $t''$ and passing trough the point with the coordinate $x_{2m}$ at the time $t_r$. 
In accordance with the conclusions of section~\ref{1}, this means that at the time $t_r$ the time derivative  
\begin{equation*}
\frac{\partial S_A}{\partial t}=\frac{2U^A}{\hbar}
\end{equation*}
for the particle $S$ position $x_1(t)$ corresponding to the position $x_{2m}$ of the particle $R$ at the same time.  Using the relations~\eqref{eq:math:ex4} and the fact that the collision takes place in the point with the coordinate $x_0$ at the time $t_r$, for the position of the subsystem $S$ corresponding to the infinite value $S^A$, we obtain 
\begin{equation*}
x_{1m}=\frac{m_2}{m_1}x_{2m}+\Bigl(1+\frac{m_2}{m_1}\Bigr)x_0.
\end{equation*}
Since the quantity $\hbar\frac{\partial S_A}{\partial t}$ for the positions $x_1\neq x_{1m}$ as well as and the other terms in the right side of equation~\eqref{eq:math:ex3} have a microscopic order of magnitudes, following the logic of~section \ref{1}, we can conclude that the solution after normalization is 
\begin{equation*}
\rho(x,y)=\delta(x_1-x_{1m})\delta(y_1-x_{1m}).
\end{equation*}

If the value $p_{2m}$ of the momentum $p_2$ is obtained after the measurement at the time $t_r$, the time derivative $\frac{\partial S_A}{\partial t}$ at the time $t_r$ has a macroscopic value for the paths of the particle $S$ having the momentum $p_1=p_{2m}$, and 
\[\operatorname{\frac{\partial S^A}{\partial t}}(x_1)\sim\left\{
\begin{array}{@{\,}r@{\quad}l@{}}
2U^A\exp\frac{i}{\hbar}p_1x\exp\Bigl(-\frac{i}{\hbar}p_1y\Bigr) & \text{for } p_1=-p_{2m},\\ \frac{\hbar}{(t-t_r)} & \text{for } p_1\neq-p_{2m}.
\end{array}\right. \]
The Fourier transform of the quantity $\rho(x_1,y_1)\frac{\partial S^A}{\partial t}$ has the form:
\begin{alignat*}{2}
\int\limits_{-\infty}^{\infty}\int\limits_{-\infty}^{\infty}\exp\frac{i}{\hbar}p_1x_1\rho(x_1,y_1)\frac{\partial S^A}{\partial t}\exp\biggl(-\frac{i}{\hbar}q_1y_1\biggr)\,dx_1\,dy_1\approx\\\approx\frac{2U^A}{\hbar}\delta(p_1+p_{2m})\delta(q_1+p_{2m}),
\end{alignat*}
Thus, for a reduced density matrix after the normalization, we have 
\begin{equation*}
\rho(p,q)=\delta(p+p_{2m})\delta(q+p_{2m}).
\end{equation*}

\section {Conclusion}

In general, a nonlinear evolution does not conserve the norm of the trace of a density matrix. The necessity of this condition, for physical reasons, requires considering the procedure of the trace normalization as a mathematical image of a real physical process. Obviously, such a process has to take place simultaneously in a whole space. Then, the nonlinear local effect on the quantum system instantly generates the change of this system state in any remote area of space. In the opinion expressed in a number of publications (for example the papers~\cite{bib:16,bib:17}), this fact under the condition of a deterministic law of a nonlocal evolution contradicts the special relativity theory. As a solution to this contradiction, the process of nonlocal evolution is proposed to be considered random. Although such an evolution mechanism does not allow signaling using the processes like a wave function reduction, it does not solve the problem of nonlocal correlation of quantum states~\cite{bib:18,bib:19}, thereby, limiting the scope of special relativity to a macroscopic level.

This contradiction with the special relativity can be eliminated if the nonlocal property is attached to the interacting objects themselves. Really, let us consider the mathematical formalism of Feynman's quantum mechanics as the image of a real physical situation. Suppose that a wave function has a material carrier and virtual paths form real paths of this carriers like in classical continuum mechanics\footnote{ If the wave function of the particle is instantaneous localized under the position measurement, then Schr\"odinger's wave function interpretation "...as giving somehow the density of the stuff of which the world is made"~\cite{bib:3} does not create the "problem of the wave packet spreading".}. Then, the set of all virtual paths in the considered volume of space describes a mechanical motion of the corresponding set of individual particles\footnote{In addition to the specific properties of individual particles of such a continuous medium~\cite{bib:20}, its fundamental difference from the classical one is that individual particles of this continuum actually form an everywhere dense set in the volume accessible to a quantum particle (there is no empty space in it), whereas for a classical continuous medium, this property is a mathematical abstraction.}. These individual particles occupy all the considered volume, even that where the wave function is zero.

If to transform the equation~\eqref{eq:math:ex3} for the function of a measure density\footnote{We use the concept of measure, not probability, to emphasize the deterministic nature of the equation for the reduced density matrix of the part of a closed system}, we obtain  
 \begin{equation*}
 \frac{\partial\rho(x,x)}{\partial t}+div{\bf j}=\frac{\partial S^A}{\partial t}\rho(x,x),
\end{equation*}
where 
\begin{equation*}
div{\bf j}=\frac{i\hbar}{2m}\frac{\partial}{\partial x}\biggl(\frac{\partial\rho(y,x)}{\partial y}-\frac{\partial\rho(y,x)}{\partial x}\biggr)\biggl|_{y=x}
\end{equation*}
is the divergence of the measure density flux. This equation has to be added by the normalization condition  
\begin{equation*}
 \int\limits_{-\infty}^{\infty}\rho(x,x)\,dx=1.
 \end{equation*}
When the measure density flux can be neglected, as, for example in the case of reduction process, change of the probability in a volume is possible if $\frac{\partial S^A}{\partial t}\neq 0$. Thus, this change is possible without any measure carrier transfer in space, i.e. without a mechanical motion. It is the result of the internal structure transformation of a quantum particle described by the density matrix. This means that there can be no question of any contradiction with special relativity. Interaction between particles under the conditions of the EPR paradox can be considered as a local when particles are matter fields in the same volume of space. They are not a distant system actually. The last assumption makes it possible to avoid violating the principle of causality~\cite{bib:21}: indeed, we have allocated objects with respect to some specific interaction, and the definition of the spacing between them requires the exploration within the framework of the special relativity.

\vfill\eject

\end{document}